\documentclass[aps,prd,,nofootinbib,superscriptaddress,tighten,preprint]{revtex4}
\usepackage[usenames]{color}
\usepackage{epsfig}
\usepackage{amsmath,bm}
\usepackage{multirow}
\usepackage{setspace}

\begin{document}

\title{Radiative corrections to the leptonic Dirac CP-violating phase}

\author{Tommy Ohlsson}

\email{tohlsson@kth.se}

\affiliation{Department of Theoretical Physics, School of
Engineering Sciences, KTH Royal Institute of Technology, AlbaNova
University Center, 106 91 Stockholm, Sweden}

\author{He Zhang}

\email{hzhang@mpi-hd.mpg.de}

\affiliation{Max-Planck-Institut f\"{u}r Kernphysik, Saupfercheckweg
1, 69117 Heidelberg, Germany}

\author{Shun Zhou}
\email{shunzhou@kth.se}

\affiliation{Department of Theoretical Physics, School of
Engineering Sciences, KTH Royal Institute of Technology, AlbaNova
University Center, 106 91 Stockholm, Sweden}

\begin{abstract}
Since the smallest leptonic mixing angle $\theta^{}_{13}$ has been
measured to be relatively large, it is quite promising to constrain
or determine the leptonic Dirac CP-violating phase $\delta$ in
future neutrino oscillation experiments. Given some typical values
of $\delta = \pi/2$, $\pi$, and $3\pi/2$ at the low-energy scale, as
well as current experimental results of the other neutrino
parameters, we perform a systematic study of the radiative
corrections to $\delta$ by using the one-loop renormalization group
equations in the minimal supersymmetric standard model and the
universal extra-dimensional model. It turns out that $\delta$ is
rather stable against radiative corrections in both models, except
for the minimal supersymmetric standard model with a very large
value of $\tan \beta$. Both cases of Majorana and Dirac neutrinos
are discussed. In addition, we use the preliminary indication of
$\delta = (1.08^{+0.28}_{-0.31})~{\pi}$ or $\delta =
(1.67^{+0.37}_{-0.77})~{\pi}$ from the latest global-fit analyses of
data from neutrino oscillation experiments to illustrate how it will
be modified by radiative corrections.
\end{abstract}

\maketitle

\section{Introduction}

In the last two decades, our knowledge on neutrinos has been greatly
improved by a number of elegant neutrino oscillation experiments
\cite{PDG}. Now, we are convinced that neutrinos are massive, and
they can transform from one flavor to another when propagating in
vacuum or in matter. The lepton flavor mixing phenomenon can be
described by a $3\times 3$ unitary matrix $V$, namely the leptonic
mixing matrix, which is conventionally parametrized through three
mixing angles $\theta^{}_{12}$, $\theta^{}_{13}$ and
$\theta^{}_{23}$, as well as three CP-violating phases $\delta$,
$\rho$ and $\sigma$, viz.,
\begin{equation}
V = U \cdot P \equiv \left(\begin{matrix}c^{}_{12} c^{}_{13} &
s^{}_{12} c^{}_{13} & s^{}_{13} e^{-{\rm i}\delta} \cr - s^{}_{12}
c^{}_{23} - c^{}_{12} s^{}_{23} s^{}_{13} e^{{\rm i}\delta} &
c^{}_{12} c^{}_{23} - s^{}_{12} s^{}_{23} s^{}_{13} e^{{\rm
i}\delta} & s^{}_{23} c^{}_{13} \cr s^{}_{12} s^{}_{23} - c^{}_{12}
c^{}_{23} s^{}_{13} e^{{\rm i}\delta} & - c^{}_{12} s^{}_{23}
-s^{}_{12} c^{}_{23} s^{}_{13} e^{{\rm i}\delta}  & c^{}_{23}
c^{}_{13}\end{matrix}\right) \cdot P \; ,
\end{equation}
where $s^{}_{ij} \equiv \sin \theta^{}_{ij}$ and $c^{}_{ij} \equiv
\cos \theta^{}_{ij}$ for $ij = 12, 13, 23$. Note that $P = {\rm
diag}(e^{{\rm i}\rho}, e^{{\rm i}\sigma}, 1)$ is a diagonal matrix
with $\rho$ and $\sigma$ being two Majorana-type CP-violating phases
if neutrinos are Majorana particles, while $P = {\bf 1}$ if
neutrinos are Dirac particles. Current experimental data indicate
that the three leptonic mixing angles are $\theta^{}_{12} \approx
34^\circ$, $\theta^{}_{13} \approx 9^\circ$ and $\theta^{}_{23}
\approx 40^\circ$. Two independent neutrino mass-squared differences
are found to be $\Delta m^2_{21} \equiv m^2_2 - m^2_1 \approx
7.5\times 10^{-5}~{\rm eV}^2$ and $|\Delta m^2_{31}| \equiv |m^2_3 -
m^2_1| \approx 2.5\times 10^{-3}~{\rm eV}^2$. The latest global-fit
results of neutrino parameters are shown in Table I. However, we are
still unclear whether the neutrino mass ordering is normal (i.e.,
$\Delta m^2_{31}
> 0$) or inverted (i.e., $\Delta m^2_{31} < 0$), and the leptonic
Dirac CP-violating phase $\delta$ remains experimentally
unconstrained.

The recent results from Daya Bay \cite{Daya} and RENO \cite{Reno}
reactor neutrino experiments have established that $\theta^{}_{13}
\approx 9^\circ$, which is rather large. Hence, it is quite
promising to determine the leptonic Dirac CP-violating phase
$\delta$ by comparing the oscillation probabilities of neutrinos and
antineutrinos in future long-baseline neutrino oscillation
experiments \cite{Branco}. In addition, the ${\rm km}^3$-scale
neutrino telescopes (e.g., IceCube and KM3NeT) could provide us with
useful and complementary information about leptonic CP violation by
precisely measuring the flavor composition of ultrahigh-energy
astrophysical neutrinos \cite{MO}. If the Deep Core of the IceCube
detector is made denser to lower the energy threshold down to a few
${\rm GeV}$, such as the proposal PINGU \cite{PINGU}, a large amount
of atmospheric neutrino events can be collected and used to
determine the neutrino mass hierarchy and perhaps the leptonic
CP-violating phase \cite{Smirnov}. On the other hand, a lot of
neutrino mass models based on discrete flavor symmetries or
phenomenological assumptions have recently been proposed to describe
the observed leptonic mixing pattern, in particular a relatively
large $\theta^{}_{13}$. Interestingly, the leptonic CP-violating
phase $\delta$ has been predicted in some models to be rather large
(e.g., $\delta > \pi/3$) or even maximal (i.e., $\delta = \pi/2$)
\cite{CP,Chen}. In other models, leptonic CP violation is shown to
be absent, namely $\delta = 0$ or $\pi$ \cite{Zhang}. It is
worthwhile to mention that the latest global-fit analyses of
neutrino oscillation experiments yield $\delta =
(1.08^{+0.28}_{-0.31})~{\pi}$ \cite{Fogli} and $\delta =
(1.67^{+0.37}_{-0.77})~{\pi}$ \cite{Schwetz}, although the $1\sigma$
errors are still quite large.\footnote{The best-fit value is found
to be $\delta = 0.8~\pi$ for the normal mass hierarchy and $\delta =
-0.03~\pi$ for the inverted mass hierarchy by another global-fit
group \cite{Valle}. However, there is no constraint on $\delta$
within the $1\sigma$ range.} Therefore, we have already obtained
some preliminary information on the leptonic CP-violating phase
$\delta$ from the global-fit analyses.
\begin{table}
\caption{The best-fit values and 1$\sigma$ ranges of the neutrino
parameters from the latest global-fit analyses of neutrino
oscillation experiments, where the normal neutrino mass hierarchy is
assumed.}
\begin{center}
\footnotesize
\begin{tabular}{cccc}
\hline \hline
\quad \quad parameter \quad \quad & \quad \quad  \quad Ref.
\cite{Fogli} \quad \quad   \quad & \quad \quad   \quad Ref.
\cite{Schwetz} \quad \quad \quad &  \quad \quad  \quad Ref.
\cite{Valle} \quad \quad \quad
\\
\hline
\multirow{2}{*}{$\sin^2\theta_{12}$}   & $0.307$ & $0.300$ & $0.320$
\\ & $0.291 - 0.325$ & $0.287 - 0.313$ & $0.303 - 0.336$ \vspace{0.2cm}
\\
\multirow{2}{*}{$\sin^2\theta_{13}$}   & $0.0241$ & $0.0230$ &
$0.0246$
\\ & $0.0216 - 0.0266$ & $0.0207 - 0.0253$ & $0.0218 - 0.0275$
\vspace{0.2cm}
\\
\multirow{2}{*}{$\sin^2\theta_{23}$}   & $0.386$ & $0.410$ & $0.427$
\\ & $0.365 - 0.410$ & $0.385 - 0.447$ & $0.400 - 0.461$ \vspace{0.2cm}
\\
\multirow{2}{*}{$\Delta m^2_{21}/10^{-5}~{\rm eV}^2$}   & $7.54$ &
$7.50$ & $7.62$
\\ & $7.32 - 7.80$ & $7.32 - 7.69$ & $7.43 - 7.81$ \vspace{0.2cm}
\\
\multirow{2}{*}{$\Delta m^2_{31}/10^{-3}~{\rm eV}^2$} & $2.51$ &
$2.47$ & $2.55$
\\ & $2.41 - 2.57$ & $2.40 - 2.54$ & $2.46 - 2.61$ \vspace{0.2cm}
\\
\multirow{2}{*}{$\delta/\pi$} & $1.08$ & $1.67$ & $0.8$
\\ & $0.77 - 1.36$ & $0.90 - 2.03$ & $0 - 2.0$ \vspace{0.2cm} \\
\hline \hline
\end{tabular}
\end{center}
\end{table}

In this work, we are concerned with how the theoretical predictions
or the observed value of $\delta$ will be modified by the radiative
corrections when running from a low-energy scale to a
superhigh-energy scale. This question {\it does} make sense if we
believe that there exists at some superhigh-energy scale a unified
theory for flavor mixing and CP violation in both quark and lepton
sectors. Once the leptonic CP-violating phase $\delta$ is measured
in future neutrino oscillation experiments, the renormalization
group (RG) evolution of $\delta$ will tell us how large or small it
will be at a given superhigh-energy scale. As a matter of fact, the
running of leptonic mixing parameters has been extensively discussed
in the literature \cite{RGE1,RGE2}, and more recently in Ref.
\cite{Luo}, where the authors concentrate on the newly measured
$\theta^{}_{13}$. Different from the previous works, we focus on
$\delta$ and perform a systematic study of its running behavior in
the minimal supersymmetric standard model (MSSM) and in the
universal extra-dimensional model (UEDM). The motivation for such a
study is two-fold: (1) The leptonic CP-violating phase $\delta$ is
the last fundamental parameter (except for the neutrino mass
hierarchy) to be measured in the future neutrino oscillation
experiments, and now both the theoretical models and the global-fit
analysis can provide us with preferred values of $\delta$ at the
low-energy scale. (2) The models with supersymmetry or extra spatial
dimensions are the most natural extensions of the SM, which can
solve the gauge hierarchy problem and offer good candidates for the
dark matter.

In lack of a complete theory for neutrino mass generation, we
implement the dimension-five Weinberg operator to account for tiny
Majorana neutrino masses \cite{Weinberg}. The RG running of $\delta$
in the case of Dirac neutrinos will be considered as well for
comparison and completeness.

The remaining part of the present paper is organized as follows. In
Sec.~II, we set up the basic framework for the RG running of
leptonic mixing parameters in the case of Majorana neutrinos. The
renormalization group equation (RGE) of $\delta$ is derived
analytically, and solved numerically. Section III is devoted to the
RG running of $\delta$ in the case of Dirac neutrinos in the MSSM.
We summarize our conclusions in Sec.~IV. The complete set of RGE's
in the SM, MSSM, and UEDM for Majorana neutrinos are collected in
Appendix A, while those in the SM and MSSM for Dirac neutrinos in
Appendix B.

\section{Running of CP-violating Phase: Majorana Neutrinos}

First of all, we derive the RGE for the leptonic CP-violating phase
$\delta$, assuming that neutrinos are Majorana particles. Without
loss of generality, we introduce the dimension-five Weinberg
operator responsible for neutrino masses \cite{Weinberg}:
\begin{equation}
-{\cal L}^{}_\nu = \frac{1}{2} \left(\overline{\ell} H\right) \cdot
\kappa \cdot \left(H^T \ell^C\right) + {\rm h.c.},
\end{equation}
where $\ell$ and $H$ stand for the lepton and Higgs doublet fields,
respectively, and $\kappa$ is a symmetric and complex matrix of the
inverse mass dimension. After electroweak symmetry breaking, the
mass matrix of three light Majorana neutrinos is given by $M^{}_\nu
= \kappa v^2$ with $v \approx 174~{\rm GeV}$ being the vacuum
expectation value (vev) of the SM Higgs field, or by $M^{}_\nu =
\kappa (v\sin\beta)^2$ with $\tan \beta$ being the ratio of the
vev's of two Higgs doublets in the MSSM. Note that we are working
within an effective theory, and consider the running of neutrino
mixing parameters below the cutoff scale $\Lambda$ where new physics
takes effects.

At one-loop level, the evolution of $\kappa$ is governed by
\cite{RGE1,RGE2}
\begin{equation}
16\pi^2 \frac{{\rm d}\kappa}{{\rm d}t} = \alpha^{}_\kappa +
C^{}_\kappa \left[\left(Y^{}_l Y^\dagger_l\right)\kappa + \kappa
\left(Y^{}_l Y^\dagger_l\right)^T\right] \; ,
\end{equation}
where $t \equiv \ln(\mu/\Lambda^{}_{\rm EW})$ with $\mu$ being an
arbitrary renormalization scale between the electroweak scale
$\Lambda^{}_{\rm EW} \approx 100~{\rm GeV}$ and a cutoff scale where
new physics comes into play, and $Y^{}_l$ is the Yukawa coupling
matrix of the charged leptons. The coefficients $\alpha^{}_\kappa$
and $C^{}_\kappa$ are flavor universal, and have been explicitly
given in Appendix A for the SM, the MSSM, and the UEDM. It is worth
stressing that Eq.~(3) takes on the same form in all the models
under consideration. However, the coefficients in the RGE's may
differ. We will distinguish them by adding the corresponding
superscripts to these coefficients, as shown in Appendix A.

\subsection{Analytical Results}
\begin{table}
\caption{Explicit expressions of ${\rm Re}\left[(U^\dagger
\dot{U})^{}_{ij}\right]$ and ${\rm Im}\left[(U^\dagger
\dot{U})^{}_{ij}\right]$ for $i \leq j$ in the standard
parametrization of leptonic mixing matrix.}
\begin{center}
\begin{tabular}{ccccc}
\hline \hline
$ij$ &~~~~~& ${\rm Re}\left[(U^\dagger \dot{U})^{}_{ij}\right]$
&~~~~~& ${\rm Im}\left[(U^\dagger \dot{U})^{}_{ij}\right]$
\\ \hline
$11$ &~~~~~& $0$ &~~~~~& $+2s^{}_{12} c^{}_{12} s^{}_{13}
s^{}_\delta \dot{\theta}^{}_{23} + c^2_{12} s^2_{13} \dot{\delta}$
\vspace{0.2cm} \\
$22$ &~~~~~& $0$ &~~~~~& $-2s^{}_{12} c^{}_{12} s^{}_{13}
s^{}_\delta \dot{\theta}^{}_{23} + s^2_{12} s^2_{13} \dot{\delta}$
\vspace{0.2cm} \\
$33$ &~~~~~& $0$ &~~~~~& $-s^2_{13} \dot{\delta}$
\vspace{0.2cm} \\
$12$ &~~~~~& $\dot{\theta}^{}_{12} + s^{}_{13} s^{}_\delta
\dot{\theta}_{23}$ &~~~~~& $-(c^2_{12} - s^2_{12})s^{}_{13}
s^{}_\delta \dot{\theta}^{}_{23} + s^{}_{12}c^{}_{12} s^2_{13}
\dot{\delta}$
\vspace{0.2cm} \\
$13$ &~~~~~& $-s^{}_{12} c^{}_{13} \dot{\theta}^{}_{23} + c^{}_{12}
c^{}_\delta \dot{\theta}^{}_{13} - c^{}_{12} s^{}_{13} c^{}_{13}
s^{}_\delta \dot{\delta}$ &~~~~~& $-c^{}_{12} s^{}_\delta
\dot{\theta}^{}_{13} - c^{}_{12} s^{}_{13} c^{}_{13} c^{}_\delta
\dot{\delta}$
\vspace{0.2cm} \\
$23$ &~~~~~& $+c^{}_{12} c^{}_{13} \dot{\theta}^{}_{23} + s^{}_{12}
c^{}_\delta \dot{\theta}_{13} - s^{}_{12} s^{}_{13} c^{}_{13}
s^{}_\delta \dot{\delta}$ &~~~~~& $-s^{}_{12} s^{}_\delta
\dot{\theta}^{}_{13} - s^{}_{12} s^{}_{13} c^{}_{13} c^{}_\delta
\dot{\delta}$ \vspace{0.2cm}
\\
\hline \hline
\end{tabular}
\end{center}
\end{table}
\begin{table}
\caption{The coefficients ${\cal R}^\alpha_{ij}$ and ${\cal
I}^\alpha_{ij}$ for $\alpha = e, \mu, \tau$ and $ij = 12, 13, 23$ in
the standard parametrization of leptonic mixing matrix.}
\begin{center}
\begin{tabular}{cccc}
\hline \hline
${\cal R}^\alpha_{ij}$ & $12$ & $13$ & $23$  \\ \hline
$e$ & $s^{}_{12} c^{}_{12} c^2_{13}$ & $c^{}_{12} s^{}_{13}
c^{}_{13}$ & $s^{}_{12} s^{}_{13} c^{}_{13}$ \vspace{0.2cm}
\\
\multirow{2}{*}{$\mu$} & $s^{}_{12} c^{}_{12} (s^2_{23} s^2_{13} -
c^2_{23})$ & ~~ $-(s^{}_{12} c^{}_{23} + c^{}_{12} s^{}_{23}
s^{}_{13} c^{}_\delta)s^{}_{23} c^{}_{13}$ & ~~~ $+(c^{}_{12}
c^{}_{23} - s^{}_{12} s^{}_{23} s^{}_{13} c^{}_\delta) s^{}_{23} c^{}_{13}$ \\
& $- (c^2_{12}-s^2_{12})s^{}_{23} c^{}_{23} s^{}_{13} c^{}_\delta$ &
~ & ~  \vspace{0.2cm} \\
\multirow{2}{*}{$\tau$} & $s^{}_{12} c^{}_{12} (c^2_{23}
s^2_{13}-s^2_{23})$ &
$+(s^{}_{12}s^{}_{23}-c^{}_{12}c^{}_{23}s^{}_{13}
c^{}_\delta)c^{}_{23} c^{}_{13}$ & $-(c^{}_{12} s^{}_{23} +
s^{}_{12} c^{}_{23} s^{}_{13} c^{}_\delta) c^{}_{23} c^{}_{13}$ \\
& $+(c^2_{12} -s^2_{12})s^{}_{23} c^{}_{23} s^{}_{13} c^{}_\delta$ &
~ & ~ \vspace{0.2cm} \\
\hline \hline \\ \hline \hline
${\cal I}^\alpha_{ij}$ & $12$ & $13$ & $23$  \\
\hline
$e$ & $0$ & $0$ & $0$ \vspace{0.2cm}
\\
$\mu$ & $+s^{}_{23} c^{}_{23} s^{}_{13} s^{}_\delta$ & $c^{}_{12}
s^2_{23} s^{}_{13} c^{}_{13} s^{}_\delta$ & $s^{}_{12} s^2_{23}
s^{}_{13} c^{}_{13} s^{}_\delta$ \vspace{0.2cm}
\\
$\tau$ & $-s^{}_{23} c^{}_{23} s^{}_{13} s^{}_\delta$ & $c^{}_{12}
c^2_{23} s^{}_{13} c^{}_{13} s^{}_\delta$ & $s^{}_{12} c^2_{23}
s^{}_{13} c^{}_{13} s^{}_\delta$ \vspace{0.2cm}
\\
\hline \hline
\end{tabular}
\end{center}
\end{table}
Since the RGE's of neutrino mass matrix $M^{}_\nu = \kappa v^2$ in
the SM and UEDM, or $M^{}_\nu = \kappa (v\sin \beta)^2$ in the MSSM,
are given by the same formula in Eq.~(3), the evolution of neutrino
mass eigenvalues and leptonic mixing parameters can be figured out
in the same way. In flavor basis, where the Yukawa coupling matrix
of the charged leptons is diagonal, namely $Y^{}_l = D^{}_l \equiv
{\rm diag}(y^{}_e, y^{}_\mu, y^{}_\tau)$, $\kappa$ can be
diagonalized by the leptonic mixing matrix $V$, namely $V^\dagger
\kappa V^* = \hat{\kappa} \equiv {\rm diag}(\kappa^{}_1,
\kappa^{}_2, \kappa^{}_3)$. Generally speaking, an arbitrary
$3\times 3$ unitary matrix $V^\prime$ can be factorized as $V^\prime
= Q U P$, where $Q = {\rm diag}(e^{{\rm i}\phi^{}_e,}, e^{{\rm
i}\phi^{}_\mu}, e^{{\rm i}\phi^{}_\tau})$ and $P = {\rm
diag}(e^{{\rm i}\rho}, e^{{\rm i}\sigma}, 1)$ are pure phase
matrices, while the unitary matrix $U$ consists of three mixing
angles $\theta^{}_{12}$, $\theta^{}_{13}$, $\theta^{}_{23}$ and the
Dirac CP-violating phase $\delta$ [cf. Eq.~(1)]. Although the phases
$\phi^{}_\alpha$ (for $\alpha = e, \mu, \tau$) are unphysical and
can be removed by rephasing the charged-lepton fields, we will keep
them in the derivation of the RGE's for neutrino masses and leptonic
mixing parameters.

Since $y^2_e \ll y^2_\mu \ll y^2_\tau$, we take into account the
dominant contribution from the tau-lepton Yukawa coupling to the RGE
of $\kappa$. Following Ref. \cite{Xing06}, one obtains
\begin{equation}
16\pi^2 \frac{{\rm d}\kappa^{}_i}{{\rm d}t} = \kappa^{}_i
\left(\alpha^{}_\kappa + 2 C^{}_\kappa y^2_\tau |U^{}_{\tau
i}|^2\right) \; ,
\end{equation}
where $\alpha^{}_\kappa$ and $C^{}_\kappa$ should bear the
corresponding superscripts when Eq.~(4) is applied to a specific
model. Given $m^{}_i = \kappa^{}_i v^2$ (for $i = 1, 2, 3$), we
observe that Eq.~(4) determines the evolution of absolute neutrino
masses. Moreover, it is straightforward to find that $U^{}_{\alpha
i}$, $\rho$, $\sigma$, and $\phi^{}_\alpha$ (for $\alpha = e, \mu,
\tau$ and $i = 1, 2, 3$) have to fulfill the following equations:
\begin{eqnarray}
&& {\rm Im}\left[(U^\dagger \dot{U})^{}_{11}\right] + \sum_\alpha
|U^{}_{\alpha 1}|^2 \dot{\phi}^{}_\alpha + \dot{\rho} =
0 \; , \nonumber \\
&& {\rm Im}\left[(U^\dagger \dot{U})^{}_{22}\right] + \sum_\alpha
|U^{}_{\alpha 2}|^2 \dot{\phi}^{}_\alpha + \dot{\sigma} = 0 \; ,
\nonumber \\
&& {\rm Im}\left[(U^\dagger \dot{U})^{}_{33}\right] + \sum_\alpha
|U^{}_{\alpha 3}|^2 \dot{\phi}^{}_\alpha  = 0 \; ,
\end{eqnarray}
and
\begin{eqnarray}
{\rm Re}\left[(U^\dagger \dot{U})^{}_{12}\right] - \sum_\alpha {\cal
I}^\alpha_{12} \dot{\phi}^{}_\alpha &=& -\frac{C^{}_\kappa
y^2_\tau}{32\pi^2} \left\{\hat{\zeta}^{}_{12}
\left[s^{}_{2(\rho-\sigma)}{\cal I}^\tau_{12} +
c^{}_{2(\rho-\sigma)} {\cal R}^\tau_{12}\right] +
\tilde{\zeta}^{}_{12} {\cal R}^\tau_{12}\right\} \; , \nonumber \\
{\rm Im}\left[(U^\dagger \dot{U})^{}_{12}\right] + \sum_\alpha {\cal
R}^\alpha_{12} \dot{\phi}^{}_\alpha &=& -\frac{C^{}_\kappa
y^2_\tau}{32\pi^2} \left\{\hat{\zeta}^{}_{12}
\left[s^{}_{2(\rho-\sigma)}{\cal R}^\tau_{12} -
c^{}_{2(\rho-\sigma)} {\cal I}^\tau_{12}\right] +
\tilde{\zeta}^{}_{12} {\cal I}^\tau_{12}\right\} \; ,  \nonumber \\
{\rm Re}\left[(U^\dagger \dot{U})^{}_{13}\right] - \sum_\alpha {\cal
I}^\alpha_{13} \dot{\phi}^{}_\alpha &=& -\frac{C^{}_\kappa
y^2_\tau}{32\pi^2} \left\{\hat{\zeta}^{}_{13}
\left[s^{}_{2\rho}{\cal I}^\tau_{13} + c^{}_{2\rho} {\cal
R}^\tau_{13}\right] +
\tilde{\zeta}^{}_{13} {\cal R}^\tau_{13}\right\} \; , \nonumber \\
{\rm Im}\left[(U^\dagger \dot{U})^{}_{13}\right] + \sum_\alpha {\cal
R}^\alpha_{13} \dot{\phi}^{}_\alpha &=& -\frac{C^{}_\kappa
y^2_\tau}{32\pi^2} \left\{\hat{\zeta}^{}_{12}
\left[s^{}_{2\rho}{\cal R}^\tau_{13} - c^{}_{2\rho} {\cal
I}^\tau_{13}\right] +
\tilde{\zeta}^{}_{13} {\cal I}^\tau_{13}\right\} \; ,  \nonumber \\
{\rm Re}\left[(U^\dagger \dot{U})^{}_{23}\right] - \sum_\alpha {\cal
I}^\alpha_{23} \dot{\phi}^{}_\alpha &=& -\frac{C^{}_\kappa
y^2_\tau}{32\pi^2} \left\{\hat{\zeta}^{}_{23}
\left[s^{}_{2\sigma}{\cal I}^\tau_{23} + c^{}_{2\sigma} {\cal
R}^\tau_{23}\right] +
\tilde{\zeta}^{}_{23} {\cal R}^\tau_{23}\right\} \; , \nonumber \\
{\rm Im}\left[(U^\dagger \dot{U})^{}_{23}\right] + \sum_\alpha {\cal
R}^\alpha_{23} \dot{\phi}^{}_\alpha &=& -\frac{C^{}_\kappa
y^2_\tau}{32\pi^2} \left\{\hat{\zeta}^{}_{23}
\left[s^{}_{2\sigma}{\cal R}^\tau_{23} - c^{}_{2\sigma} {\cal
I}^\tau_{23}\right] + \tilde{\zeta}^{}_{23} {\cal
I}^\tau_{23}\right\} \; ,
\end{eqnarray}
where $\hat{\zeta}^{}_{ij} \equiv 4\kappa^{}_i
\kappa^{}_j/(\kappa^2_i - \kappa^2_j)$ and $\tilde{\zeta}^{}_{ij}
\equiv  2(\kappa^2_i + \kappa^2_j)/(\kappa^2_i - \kappa^2_j)$ have
been defined, and the overdot refers to the derivative with respect
to the running parameter $t$. In addition, ${\cal R}^\alpha_{ij}
\equiv {\rm Re}\left(U^*_{\alpha i} U^{}_{\alpha j}\right)$ and
${\cal I}^\alpha_{ij} \equiv {\rm Im}\left(U^*_{\alpha i}
U^{}_{\alpha j}\right)$. Given the standard parametrization of $U$
in Eq.~(1), the matrix elements of $U^\dagger \dot{U}$ are shown in
Table II, while the coefficients ${\cal R}^\alpha_{ij}$ and ${\cal
I}^\alpha_{ij}$ are given in Table III. Note that Eqs.~(5) and (6)
form an array of differential equations linear in
$\{\dot{\theta}^{}_{12}, \dot{\theta}^{}_{13}, \dot{\theta}^{}_{23},
\dot{\delta}, \dot{\rho}, \dot{\sigma}, \dot{\phi}^{}_e,
\dot{\phi}^{}_\mu, \dot{\phi}^{}_\tau\}$, which can be explicitly
solved. As a result, the RGE of $\delta$ can be approximately
written as
\begin{eqnarray}
\dot{\delta} &\approx& \frac{C^{}_\kappa y^2_\tau}{32\pi^2}
\left\{\frac{s^{}_{12} c^{}_{12} s^{}_{23} c^{}_{23}}{s^{}_{13}}
\left[s^{}_\delta (\tilde{\zeta}^{}_{32} - \tilde{\zeta}^{}_{31}) +
(s^{}_{(\delta + 2\sigma)} \hat{\zeta}^{}_{32} -
s^{}_{(\delta+2\rho)} \hat{\zeta}^{}_{31})\right] \right. \nonumber \\
&~&  - \hat{\zeta}^{}_{21} s^2_{23} s^{}_{2(\rho-\sigma)}- (c^2_{23}
- s^2_{23}) (s^{}_{2\rho} s^2_{12} \hat{\zeta}^{}_{31} +
s^{}_{2\sigma} c^2_{12} \hat{\zeta}^{}_{32}) + c^2_{23}
(s^{}_{2(\delta + \rho)} c^2_{12} \hat{\zeta}^{}_{31} +
s^{}_{2(\delta + \sigma)} s^2_{12} \hat{\zeta}^{}_{32}) ~~~
\nonumber \\
&~& - \frac{s^{}_{23} c^{}_{23} s^{}_{13}}{s^{}_{12} c^{}_{12}}
\left[\tilde{\zeta}^{}_{21} s^{}_\delta - \hat{\zeta}^{}_{21}
(s^{}_{(\delta+2\rho-2\sigma)} c^2_{12} + s^{}_{(\delta -2 \rho +
2\sigma)} s^2_{12})\right]\left. + s^2_{13} c^2_{23}
s^{}_{2(\rho-\sigma)}\hat{\zeta}^{}_{21} \right\} \; .
\end{eqnarray}
Since the last two terms in the third line of Eq.~(7) are
proportional to $s^{}_{13}$ and $s^2_{13}$, we have neglected the
terms further suppressed by ${\cal O}(\Delta m^2_{21}/|\Delta
m^2_{31}|)$. If neutrino masses are nearly degenerate $m^2_i \gg
|\Delta m^2_{31}| \gg \Delta m^2_{21}$, which will always be assumed
in the following, we have $\hat{\zeta}^{}_{ij} \approx
\tilde{\zeta}^{}_{ij} \approx 4m^2_i/(m^2_i - m^2_j)$ and
$\hat{\zeta}^{}_{21} \gg |\hat{\zeta}^{}_{32}|,
|\hat{\zeta}^{}_{31}| \gg 1$, and thus, the RG evolution of $\delta$
could be significant. To next-to-leading order, Eq.~(7) approximates
to
\begin{equation}
\dot{\delta} \approx -\frac{C^{}_\kappa
y^2_\tau}{8\pi^2}\frac{m^2_1}{\Delta m^2_{21}} \left\{s^2_{23}
s^{}_{2(\rho - \sigma)} + \frac{2s^{}_{23}c^{}_{23}}{s^{}_{12}
c^{}_{12} s^{}_{13}} \left[ s^2_{13} c^{}_{(\delta + \rho - \sigma)}
+ \frac{\Delta m^2_{21}}{\Delta m^2_{31}} s^2_{12}c^2_{12}
c^{}_{(\delta+\rho+\sigma)}s^{}_{(\rho-\sigma)} \right]\right\} \; ,
\end{equation}
where we have taken $m^{}_1$ as the absolute neutrino mass and
ignored the difference between $\Delta m^2_{31}$ and $\Delta
m^2_{32}$. Some comments are in order:
\begin{itemize}
\item In general, the evolution of $\delta$ is
dominated by the leading-order term $-\hat{\zeta}^{}_{21} s^2_{23}
s^{}_{2(\rho - \sigma)}$ on the right-hand side of Eq.~(7). At
higher order, if the terms suppressed by
$|\hat{\zeta}^{}_{31}|/\hat{\zeta}^{}_{21} = \Delta m^2_{21}/|\Delta
m^2_{31}| \approx 1/30$ are taken into account, then those by
$s^2_{13} \approx 1/40$ should also be kept for consistency, since
they are of the same order of magnitude, as we have done in Eq.~(8).
The relative error in Eq.~(7) is at the level of $s^{}_{13}
|\hat{\zeta}^{}_{31}|/\hat{\zeta}^{}_{21} \approx 0.5~\%$, given the
best-fit values of $\theta^{}_{13}$ and neutrino mass-squared
differences.

\item It is evident from Eq.~(7) that the evolution of $\delta$ is entangled
with that of three mixing angles and two Majorana CP-violating
phases. In particular, it depends crucially on the Majorana phases
$\rho$ and $\sigma$. It has been found that the Dirac CP-violating
phase $\delta$ can be radiatively generated from $\rho$ and
$\sigma$, even if the initial value of $\delta$ is vanishing
\cite{Luo05}. On the other hand, the RG evolution of $\delta$
becomes negligible when $\rho \approx \sigma$, while the mixing
angle $\theta^{}_{12}$ is quite sensitive to the RG effect in this
case.

\item The RGE's of $\delta$ in the SM, the MSSM, and the UEDM are
given by the same formula in Eq.~(7), but with different values of
the coefficient $C^{}_\kappa$. We have $C^{\rm SM}_\kappa = -3/2$ in
the SM, while $C^{\rm MSSM}_\kappa = 1$ in the MSSM and $C^{\rm
UEDM}_\kappa = -3(1+s)/2$ in the UEDM, respectively. Therefore,
given the same Majorana CP-violating phases and leptonic mixing
angles, the evolution of $\delta$ in the MSSM will be in the
direction opposite to that in the SM and the UEDM.
\end{itemize}

Finally, we observe from Eq.~(5) that the identity $\dot{\phi}^{}_e
+ \dot{\phi}^{}_\mu + \dot{\phi}^{}_\tau + \dot{\rho} + \dot{\sigma}
= 0$ holds in the standard parametrization of $U$. The proof is as
follows. Given a general non-singular matrix $X$, whose elements are
functions of the running parameter $t$, one can prove that ${\rm d}
[{\rm det}(X)]/{\rm d}t = {\rm det}(X) \cdot {\rm tr}[X^{-1} ({\rm
d}X/{\rm d}t)]$. If we take $X$ to be a unitary matrix $U$ with
${\rm det}(U) = 1$ and $U^{-1} = U^\dagger$, then ${\rm
tr}(U^\dagger \dot{U}) = 0$ can be obtained. This observation
together with Eq.~(5) leads to the identity $\dot{\phi}^{}_e +
\dot{\phi}^{}_\mu + \dot{\phi}^{}_\tau + \dot{\rho} + \dot{\sigma} =
0$. However, this identity depends on the specific parametrization
of $U$. For instance, if ${\rm det}(U) = e^{-{\rm i}\phi}$ with
$\phi$ being the Dirac CP-violating phase, then we have $\dot{\phi}
= \dot{\phi}^{}_e + \dot{\phi}^{}_\mu + \dot{\phi}^{}_\tau +
\dot{\rho} + \dot{\sigma}$, as shown in Ref. \cite{Xing06}.

\subsection{Numerical Results}

We proceed in this subsection with the numerical solution to the RGE
of the leptonic Dirac CP-violating phase $\delta$. Since the
evolution of $\delta$ in the SM is negligible even in the case of a
nearly-degenerate neutrino mass spectrum, we consider only the MSSM
and the UEDM. Note that no approximations to the RGE of $\delta$
will be made in our numerical calculations. Our numerical results
are shown in Fig. 1, and the main points are summarized as follows.

In the MSSM, we have taken two typical values of $\tan \beta = 10$
and $\tan \beta = 30$ for illustration. In both cases, the absolute
neutrino mass $m^{}_1 = 0.1~{\rm eV}$ is assumed, which is
consistent with the cosmological bound $m^{}_1 + m^{}_2 + m^{}_3 <
1.3~{\rm eV}$ ($95~\%$ C.L.) from the WMAP Collaboration
\cite{bound}. For the initial values of $\delta$ at the electroweak
scale, we have chosen $\delta = \pi/2$, $\pi$, and $3\pi/2$ as
typical examples. Since the tau-lepton Yukawa coupling is given by
$y^2_\tau = m^2_\tau (1+\tan^2 \beta)/v^2$ in the MSSM, the
evolution of $\delta$ should be significantly enhanced for a large
value of $\tan \beta$, as shown in the upper plots of Fig.~1. For
$\tan \beta = 30$, the RG running of $\delta$ is quite significant.
In particular, even if $\delta = \pi$ is found at the low-energy
scale, namely, there is no CP-violating effect in neutrino
oscillation experiments, the maximal CP-violating phase $\delta =
\pi/2$ or $3\pi/2$ can be achieved at the cutoff scale $\Lambda =
10^{14}~{\rm GeV}$. In other words, one can change from the scenario
with a zero CP-violating phase to that with a maximal CP-violating
phase, or vice versa. For $\tan \beta = 10$, the radiative
correction to $\delta$ is at most $10~\%$ even at $\Lambda =
10^{14}~{\rm GeV}$.

In the UEDM, we have input two different values of the absolute
neutrino mass $m^{}_1 = 0.1~{\rm eV}$ and $m^{}_1 = 0.5~{\rm eV}$.
As shown in the lower plots of Fig.~1, $\delta$ is rather stable
against radiative corrections for $m^{}_1 = 0.1~{\rm eV}$. Even for
$m^{}_1 = 0.5~{\rm eV}$, which is marginally in tension with the
cosmological bound, the relative change of $\delta$ at the cutoff
scale $\Lambda = 3\times 10^4~{\rm GeV}$ is not larger than $10~\%$.
The cutoff scale $\Lambda = 3\times 10^4~{\rm GeV}$ in the UEDM has
been chosen to avoid the Landau pole, where the Higgs mass is $M_H =
125~{\rm GeV}$ and $R^{-1} = 10~{\rm TeV}$ with $R$ being the radius
of the compactified extra dimension. Since the valid energy range in
the UEDM is much smaller than that in the MSSM, the RG running does
not develop as much. However, it should be noted that the RG running
in UEDM is actually in the form of a power law, and thus can be more
significant than in the SM and in the MSSM.

\begin{figure}[t]
\includegraphics[width=0.9\textwidth]{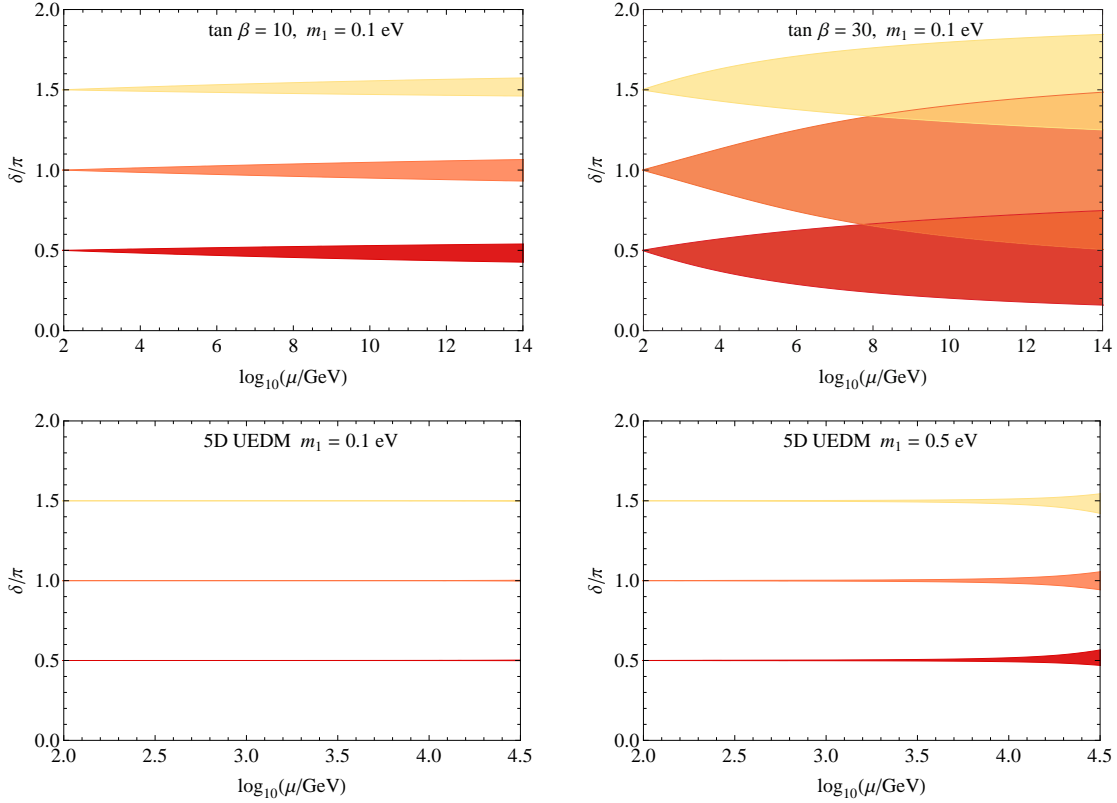}
\caption{Evolution of $\delta$ for Majorana neutrinos in the MSSM
(upper plots) and in the UEDM (lower plots). The initial values
$\delta=\pi/2$, $\delta=\pi$, and $\delta=3\pi/2$ are assumed, while
the Majorana CP-violating phases $\rho$ and $\sigma$ are
marginalized. The values of $\theta^{}_{12}$, $\theta^{}_{13}$,
$\theta^{}_{23}$ and $\Delta m^2_{21}$, $\Delta m^2_{31}$ in the
$1\sigma$ ranges from the global-fit analysis (for $\Delta m^2_{31}
> 0$) have been used as input \cite{Schwetz}.}
\end{figure}

It should also be noted that the Majorana CP-violating phases $\rho$
and $\sigma$ have been marginalized over the range $[0, \pi)$ in our
numerical results. If the specific values of $\rho$ and $\sigma$ are
chosen, the variation of $\delta$ will be even smaller. Therefore,
we conclude that the leptonic Dirac CP-violating phase $\delta$ is
stable against radiative corrections in all the models under
consideration, except for the MSSM with a large value of $\tan
\beta$. In comparison, the Dirac CP-violating phase in the quark
sector is stable even in the MSSM with a large value of $\tan
\beta$, since the quark mass spectrum is strongly hierarchical.

\begin{figure}[t]
\includegraphics[width=0.9\textwidth]{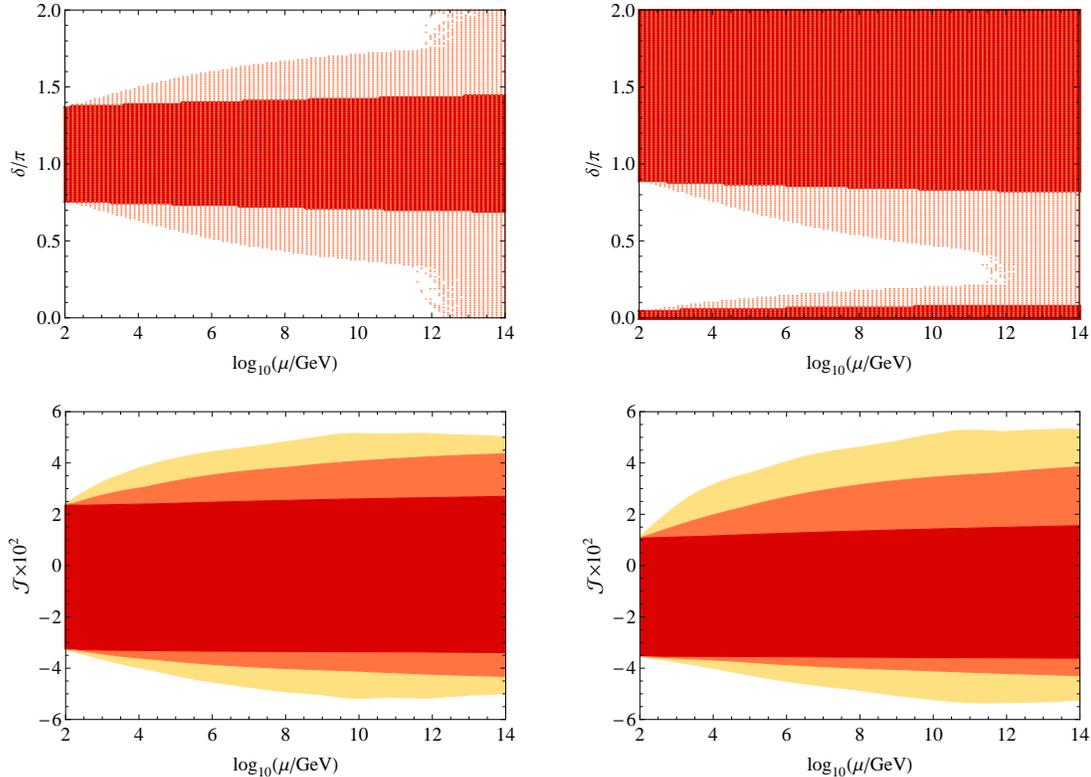}\vspace{-0.cm}
\caption{Allowed values of the leptonic CP-violating phase $\delta$
(upper plots) and the Jarlskog invariant ${\cal J}$ (lower plots)
for Majorana neutrinos at $1\sigma$ C.L. with $\tan\beta=10$ (dark
red or dark gray) and $\tan\beta=30$ (light red or gray) in the
MSSM. The result of ${\cal J}$ in the MSSM with $\tan \beta = 50$ is
also given in the lower plots (yellow or light gray). The global-fit
data from Ref.~\cite{Fogli} are adopted for the left column, while
that from Ref.~\cite{Schwetz} for the right column.}
\end{figure}

Now, we turn to the RG running behavior of $\delta$ by taking the
global-fit results $\delta = (1.08^{+0.28}_{-0.31})~{\pi}$
\cite{Fogli} and $\delta = (1.67^{+0.37}_{-0.77})~{\pi}$
\cite{Schwetz} as input. Since the present uncertainty is large, we
will choose the $1\sigma$ range for illustration. In the upper plots
of Fig.~2, the allowed regions of $\delta$ at the superhigh-energy
scale have been given in the MSSM. In the case of $\tan \beta = 30$,
one can observe that $\delta$ is almost arbitrary within $[0, 2\pi)$
due to the large uncertainty of the input, so any predictions for
$\delta$ from a high-energy flavor model could be made consistent
with the low-energy observations by the RG running. This is true for
the global-fit results from both groups \cite{Fogli,Schwetz}. In
reality, any observable effects of CP violation should be related to
the Jarlskog invariant ${\cal J} \equiv s^{}_{12} c^{}_{12}
s^{}_{23} c^{}_{23} s^{}_{13} c^2_{13} s^{}_\delta$. Therefore, we
also show the RG running of ${\cal J}$ in the MSSM for $\tan \beta =
10$, $30$, $50$, in the lower plots of Fig.~2. It can be observed
that ${\cal J}$ at a superhigh-energy scale could be quite different
from that at the low-energy scale, in particular for $\tan \beta =
30$ and $\tan \beta = 50$.

\section{Running of CP-violating Phase: Dirac Neutrinos}

The possibility for neutrinos to be Dirac particles has never been
experimentally excluded. Moreover, it has been shown that the
leptogenesis mechanism responsible for the matter-antimatter
asymmetry in our Universe also works well in a different way for
Dirac neutrinos \cite{Dlept}. Hence, we assume neutrinos to be Dirac
particles, and give them masses through the coupling to the Higgs
doublet $- \overline{\ell}^{}_{\rm L} Y^{}_\nu \nu^{}_{\rm R} H +
{\rm h.c.}$ with $Y^{}_\nu$ being the neutrino Yukawa coupling
matrix. It is convenient to write the RGE's of Dirac neutrino
parameters as \cite{LindnerD}
\begin{equation}
16\pi^2 \frac{{\rm d}\omega}{{\rm d}t} = 2\alpha^{}_\nu \omega +
C^{}_{\nu, l} \left[\left(Y^{}_l Y^\dagger_l\right) \omega + \omega
(Y^{}_l Y^\dagger_l)\right] \; ,
\end{equation}
where $\omega \equiv Y^{}_\nu Y^\dagger_\nu$ has been defined. The
RGE's of $\kappa$ in the SM and the MSSM take the same form in
Eq.~(8), but with different coefficients $\alpha^{}_\nu$ and
$C^{}_{\nu, l}$, as given in Appendix B. Since the beta function for
Dirac neutrino Yukawa couplings is currently not available in the
UEDM, we consider only the SM and the MSSM. Similarly, as in the
Majorana neutrino case, we find the RGE for the leptonic Dirac
CP-violating phase $\delta$ in the case of Dirac neutrinos
\begin{equation}
\dot{\delta} \approx - \frac{C^{}_{\nu, l} y^2_\tau}{16\pi^2}
\frac{s^{}_{23} c^{}_{23} s^{}_{13} s^{}_\delta}{s^{}_{12}
c^{}_{12}} \left[\xi^{}_{21} + (c^2_{12} \xi^{}_{32} - s^2_{12}
\xi^{}_{31}) + \frac{s^2_{12}c^2_{12}}{s^2_{13}} (\xi^{}_{32} -
\xi^{}_{31})\right] \; ,
\end{equation}
where $\xi^{}_{ij} \equiv (m^2_i + m^2_j)/(m^2_i - m^2_j)$ has been
defined. The relative error in the above equation is at the level of
$s^{}_{13} (\Delta m^2_{21}/|\Delta m^2_{31}|)^2 \sim 10^{-4}$. It
is worth mentioning that the last term in Eq.~(10) is comparable in
magnitude to the second term, since the suppression by a factor of
$\Delta m^2_{21}/|\Delta m^2_{31}|$ is compensated by the
enhancement from $1/s^2_{13}$. Some general comments are in order:
\begin{itemize}
\item The evolution of $\delta$ is proportional to $s^{}_\delta$ at
all orders, so $\delta$ will be kept unchanged by the RG running if
$s^{}_\delta = 0$, namely, $\delta =0$ or $\delta = \pi$. In other
words, if leptonic CP violation is absent at low energies, it will
never be generated by RG running. This is quite different from the
Majorana case, where $\delta$ can be radiatively generated via the
non-vanishing Majorana CP-violating phases even if $\delta = 0$ or
$\delta = \pi$ has been used as an initial condition.

\item Two qualitative differences between the SM and the MSSM should
be noted. First, the tau-Yukawa coupling $y^2_\tau = m^2_\tau (1 +
\tan^2 \beta)/v^2$ in the MSSM is significantly enhanced for a large
value of $\tan \beta$. Hence, the RG effect is more remarkable than
that in the SM. Second, the coefficient $C^{}_{\nu, l}$ takes
opposite signs in the SM and in the MSSM, indicating the evolution
of $\delta$ in opposite directions in these two models.
\end{itemize}
\begin{figure}[t]
\includegraphics[width=0.95\textwidth]{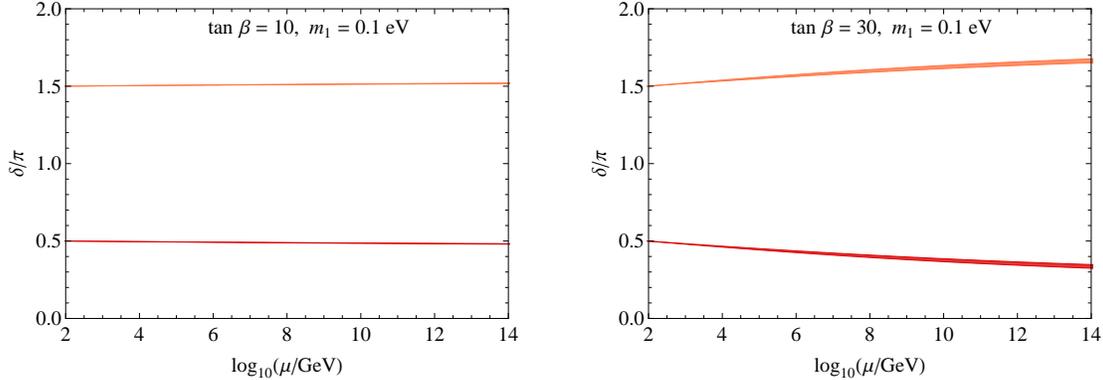}\vspace{-0.cm}
\caption{Evolution of $\delta$ for Dirac neutrinos in the MSSM for
$\tan \beta = 10$ (left plot) and $\tan \beta = 30$ (right plot).
The initial values $\delta=\pi/2$ and $\delta=3\pi/2$ are assumed,
and the values of $\theta^{}_{12}$, $\theta^{}_{13}$,
$\theta^{}_{23}$ and $\Delta m^2_{21}$, $\Delta m^2_{31}$ in the
$1\sigma$ ranges from the global-fit analysis (for $\Delta m^2_{31}
> 0$) have been used as input \cite{Schwetz}.}
\end{figure}
\begin{figure}[t]
\includegraphics[width=0.9\textwidth]{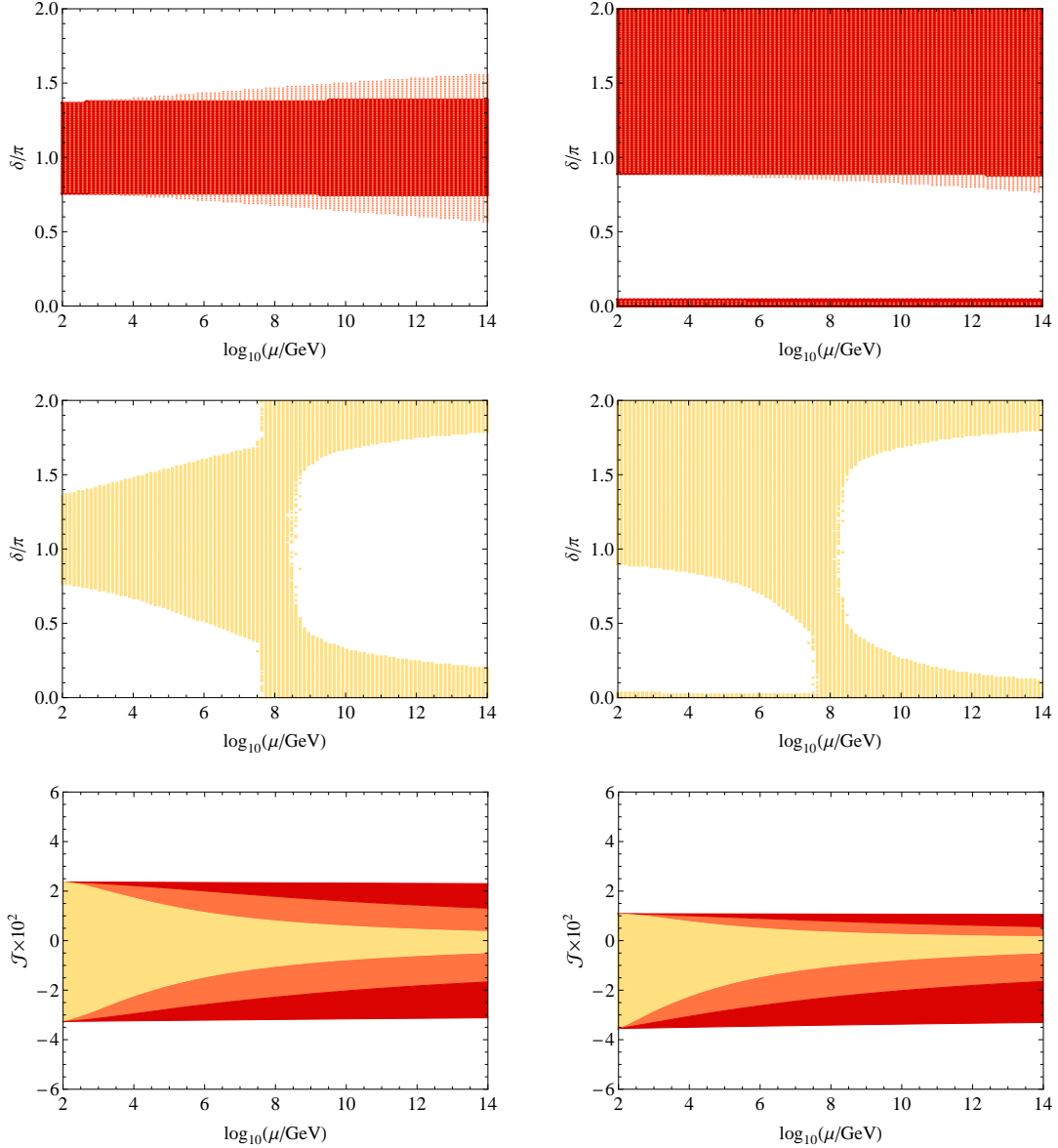}\vspace{-0.cm}
\caption{Allowed values of the leptonic Dirac CP-violating phase
$\delta$ (upper and middle plots) and the Jarlskog invariant ${\cal
J}$ (lower plots) for Dirac neutrinos at $1\sigma$ C.L. with
$\tan\beta=10$ (dark red or dark gray), $\tan\beta=30$ (light red or
gray) and $\tan \beta = 50$ (yellow or light gray) in the MSSM. The
absolute neutrino mass $m^{}_1 = 0.1~{\rm eV}$ has been assumed. The
global-fit data from Ref.~\cite{Fogli} are adopted for the left
column, while that from Ref.~\cite{Schwetz} for the right column.}
\end{figure}

To illustrate the RG running behavior of $\delta$ in the Dirac
neutrino case, we have shown in Fig.~3 two typical examples in the
MSSM. In both examples, the initial values of $\delta$ have been
taken to be $\pi/2$ and $3\pi/2$, and the absolute neutrino mass is
$m^{}_1 = 0.1~{\rm eV}$. The left plot is for $\tan \beta = 10$,
while the right for $\tan \beta = 30$. Note that the beta function
of $\delta$ is proportional to $-s^{}_\delta$ in Eq.~(9), where
$C^{}_{\nu, l} = 1$ in the MSSM. Therefore, $\delta$ increases for
$\delta = 3\pi/2$, while it decreases for $\delta = \pi/2$, as the
energy scale evolves towards higher energies. This feature can be
clearly observed in Fig.~3. Furthermore, the variation of $\delta$
at any energy scale is quite small, compared to that in the case of
Majorana neutrinos, where the arbitrary Majorana CP-violating phases
play an important role in the evolution of $\delta$. As we have
already mentioned, $\delta$ will be kept unchanged if the initial
values lead to $s^{}_\delta = 0$, so the trivial cases of $\delta =
0$ and $\delta = \pi$ have not been considered.

Now, we continue with the global-fit results of $\delta$ in
Refs.~\cite{Fogli,Schwetz} as initial values. The RG running of
$\delta$ in the MSSM for $\tan \beta = 10, 30$ and $\tan \beta = 50$
have been shown in the upper and middle plots of Fig.~4,
respectively. As before, the absolute neutrino mass $m^{}_1 =
0.1~{\rm eV}$ is assumed. In the former case, the RG running effects
are insignificant, which is in accordance with the results in
Fig.~3. In the latter case, however, it is interesting to note that
a wide range of values $\delta \in [0.2 \pi, 1.8 \pi]$ cannot be
reached at the superhigh-energy scale $\Lambda = 10^{14}~{\rm GeV}$,
no matter what initial value of $\delta$ is chosen. The reason for
this behavior is that the mixing angle $\theta^{}_{13}$ is
approaching zero around $\Lambda^\prime = 10^8~{\rm GeV}$. In the
limit of an extremely small value of $\theta^{}_{13}$, Eq.~(8) can
be written as
\begin{equation}
\dot{\delta} \approx -\frac{y^2_\tau}{16\pi^2} s^{}_{12} c^{}_{12}
s^{}_{23} c^{}_{23} s^{}_\delta s^{-1}_{13} (\xi^{}_{32} -
\xi^{}_{31}) \; ,
\end{equation}
where $C^{}_{\nu, l} = 1$ has been chosen for the MSSM. Therefore,
the RG running of $\delta$ will be rapidly accelerated around
$\Lambda^\prime = 10^8~{\rm GeV}$ to the large-value region for
$s^{}_\delta < 0$ (i.e., $\delta > \pi$), while to the small-value
region for $s^{}_\delta > 0$ (i.e., $\delta < \pi$). This
observation applies also to any initial value of $\delta$. In fact,
we have numerically checked the whole parameter region of $\delta
\in [0, 2\pi)$ at low energies, and found that only $[0, 0.2\pi]$
and $[1.8\pi, 2\pi)$ can be reached at high energies. However, the
exact allowed range of $\delta$ at high-energy scales really depends
on the initial values of $\delta$ and three mixing angles. For
$\delta = \pi$, the RG running of $\delta$ will be absent, but
$\theta^{}_{13}$ becomes negative above $\Lambda^\prime = 10^8~{\rm
GeV}$, so we have to redefine $\delta \to \delta \pm \pi$ to make
$\theta^{}_{13}$ positive, leading to $\delta = 0$ or $2\pi$ at
high-energy scales. In the lower plots of Fig. 4, the evolution of
the Jarlskog invariant ${\cal J}$ is shown. Unlike the Dirac
CP-violating phase $\delta$ itself, the physical observable ${\cal
J}$ evolves smoothly over the whole range of energy scales, as it
should. For $\tan \beta = 50$, the value of $|{\cal J}|$ can
initially be as large as $2~\%$, it becomes vanishingly small at
$\Lambda = 10^{14}~{\rm GeV}$. One reason for this is that $\delta$
shrinks into a small region around $0$ or $2\pi$ at the high-energy
scale, as indicated in the middle plots of Fig.~4. Obviously, the
evolution of the three mixing angles is also relevant here.

\section{Further Discussions}

In Secs.~II and III, we have examined the RG running behaviors of
the leptonic Dirac CP-violating phase $\delta$ in the cases of
Majorana neutrinos and Dirac neutrinos, respectively. Now, we
compare these two cases and summarize the main differences:
\begin{itemize}
\item In the Majorana case, the two Majorana CP-violating phases
are playing a crucial role in the RG running of $\delta$. One can
start from a CP-conserving scenario with $\delta = 0$ or $\pi$ at
the low-energy scale, and end up with a CP-violating scenario even
with $\delta = \pi/2$ or $3\pi/2$. In the Dirac case, the evolution
of $\delta$ is proportional to $s^{}_\delta$, so the CP conservation
at the low-energy scale definitely implies that CP violation is
absent at a superhigh-energy scale.

\item The mixing angle $\theta^{}_{13}$ could approach zero at some
high-energy scale $\Lambda^\prime$ in both cases if a large value of
$\tan \beta$ is assumed in the MSSM. On the other hand, there exist
in the RGE's of $\delta$ some terms inversely proportional to
$s^{}_{13}$. Therefore, the RG running behavior of $\delta$ will be
dramatically changed around $\Lambda^\prime$. Given the global-fit
values of $\delta$ within the $1\sigma$ range, it turns out that
$\delta$ could be arbitrary at the high-energy scale in the Majorana
case due to the marginalization over $\rho$ and $\sigma$. In the
Dirac case, $\delta$ is found to be in two narrow ranges $[0,
0.2\pi]$ or $[1.8\pi, 2\pi)$ in the MSSM with $\tan \beta = 50$.
\end{itemize}

However, if a concrete mass model for Majorana neutrinos or Dirac
neutrinos is assumed, the RG running of $\delta$ may depend on the
model details. In particular, when new particles or interactions
come into play at some intermediate energy scale, the RGE's of the
neutrino parameters are completely changed \cite{threshold}. Hence,
we have assumed that this is not the case in the previous
discussions, at least below the cutoff scale.

As we have mentioned before, many flavor symmetry models, which are
intended for describing the observed leptonic mixing angles, predict
the leptonic Dirac CP-violating phase $\delta$. For instance, it has
been shown in Ref. \cite{Chen} that $\delta \approx 2\pi/3$ (or
$4\pi/3$) and $\delta \approx \pi/3$ (or $5\pi/3$) for different
breaking patterns of the $A^{}_4$ flavor symmetry in the type-I
seesaw model, where three heavy right-handed neutrino singlets are
introduced to realize the dimension-five Weinberg operator. If the
vacuum alignment problem is further solved in the framework of
supersymmetry, significant radiative corrections to these
theoretical predictions of $\delta$ could be possible. Thus, the
leptonic Dirac CP-violating phase to be measured in neutrino
oscillation experiments is related by the RG running to the
theoretical prediction at the seesaw scale. On the other hand, the
CP-violating and out-of-equilibrium decays of the heavy right-handed
neutrinos can generate the lepton number asymmetry in the early
Universe, which will be converted into the baryon number asymmetry
via the SM sphaleron processes. In this case, the leptonic CP
violation in neutrino oscillations can be associated with the
matter-antimatter asymmetry in our Universe.

\section{Summary}

Thanks to the recent measurements of $\theta^{}_{13}$ in the Daya
Bay and RENO experiments, the discovery of CP violation in neutrino
oscillation experiments seems to be promising if the leptonic CP
violation really exists and the leptonic Dirac CP-violating phase
$\delta$ happens to be far away from $0$ or $\pi$. On the other
hand, we have already had a preliminary result for the leptonic
CP-violating phase $\delta$ from the global-fit analysis of all
kinds of neutrino oscillation experiments, namely $\delta =
(1.08^{+0.28}_{-0.31})~{\pi}$ \cite{Fogli} and $\delta =
(1.67^{+0.37}_{-0.77})~{\pi}$ \cite{Schwetz}. Therefore, we are well
motivated to study the RG running of $\delta$ from the low-energy
scale to a superhigh-energy scale, where a unified model for fermion
masses, flavor mixing, and CP violation is expected.

In the case of Majorana neutrinos, we have introduced the
dimension-five Weinberg operator to account for neutrino masses. The
RGE of $\delta$ has been derived analytically in great detail for
the SM, the MSSM, and the UEDM, and a self-consistent approximation
to it has been given as well. By a self-consistent approximation, we
mean that the RGE of $\delta$ has been expanded in terms of
$s^2_{13}$ and $\Delta m^2_{21}/|\Delta m^2_{31}|$, and all the
terms of the same order of magnitude should be preserved. It turns
out that $\delta$ is rather stable against radiative corrections in
all these models, except for the case of a large $\tan \beta$ in the
MSSM (e.g., $\tan \beta = 30$ together with a nearly degenerate
neutrino mass spectrum). In this case, the Majorana CP-violating
phases play an important role in the evolution of $\delta$ such that
a maximal phase $\delta = \pi/2$ or $3\pi/2$ can be radiatively
generated at a superhigh-energy scale even if $\delta = \pi$ (i.e.,
no CP-violating effects in neutrino oscillation experiments) at the
low-energy scale. The evolution of $\delta$ and the Jarlskog
invariant ${\cal J}$ have been illustrated by taking the $1\sigma$
global-fit results of $\delta$ as input.

In the case of Dirac neutrinos, we have derived the RGE of $\delta$
in the SM and MSSM, and the self-consistent approximation to it has
been made. Note that a nearly degenerate neutrino mass spectrum and
the absolute neutrino mass $m^{}_1 = 0.1~{\rm eV}$ are assumed in
our analysis. The RG running effect of $\delta$ can be neglected in
the SM and in the MSSM with a small $\tan \beta$ (e.g., $\tan \beta
\leq 10$). However, $\delta$ can be modified by more than $30~\%$
for $\tan \beta = 30$. The evolution of $\delta$ and the Jarlskog
invariant ${\cal J}$ have been examined by inputting the $1\sigma$
global-fit results of $\delta$. In the case of $\tan \beta = 50$,
$\delta$ in the range of $[0.2\pi, 1.8\pi]$ is found to be
unreachable at $\Lambda = 10^{14}~{\rm GeV}$, since the mixing angle
$\theta^{}_{13}$ approaches zero at some intermediate scale (e.g.,
$\Lambda^\prime = 10^8~{\rm GeV}$), which forces $\delta$ to be in a
large-value region for $\delta > \pi$ or a small-value region for
$\delta < \pi$. At the same time, the Jarlskog invariant ${\cal J}$
becomes vanishingly small at a superhigh-energy scale.

As we already know some information and will soon learn more about
the leptonic Dirac CP-violating phase $\delta$, it is thus
meaningful to see how large it will be at a superhigh-energy scale.
At such an energy scale, the leptonic Dirac CP-violating phase might
be related to the quark Dirac CP-violating phase in a unified flavor
model, or to the generation of matter-antimatter asymmetry in our
Universe via the leptogenesis mechanism. In any case, the precise
determination of $\delta$ in the ongoing and upcoming neutrino
oscillation experiments or at a future neutrino factory will shed
light on the flavor dynamics at a high-energy scale.

\acknowledgements

H.Z. would like to thank for financial support the G\"{o}ran
Gustafsson Foundation, and for the hospitality the KTH Royal
Institute of Technology, where part of this work was performed. This
work was supported by the Swedish Research Council
(Vetenskapsr{\aa}det), contract no. 621-2011-3985 (T.O.), the Max
Planck Society through the Strategic Innovation Fund in the project
MANITOP (H.Z.), and the G\"{o}ran Gustafsson Foundation (S.Z.).

\appendix

\section{RGE's for Majorana neutrinos}

\subsection{The SM}

In the SM extended with the dimension-five Weinberg operator, the
RGE for $\kappa$ has already been given in Eq.~(3), while those for
the Yukawa coupling matrices $Y^{}_{\rm f}$ of charged fermions
(i.e., ${\rm f} = l$ for charged leptons, ${\rm f} = {\rm u}$ for
up-type quarks and ${\rm f} = {\rm d}$ for down-type quarks) can be
written as
\begin{eqnarray}
16\pi^2 \frac{{\rm d}Y^{}_l}{{\rm d}t} &=& \left[\alpha^{\rm SM}_l +
C^{\rm SM}_{l,l} \left(Y^{}_l Y^\dagger_l\right)\right] Y^{}_l \; ,
\nonumber \\
16\pi^2 \frac{{\rm d}Y^{}_{\rm u}}{{\rm d}t} &=& \left[ \alpha^{\rm
SM}_{\rm u} + C^{\rm SM}_{\rm u,u} \left(Y^{}_{\rm u} Y^\dagger_{\rm
u} \right) + C^{\rm SM}_{\rm u,d} \left(Y^{}_{\rm d} Y^\dagger_{\rm
d}\right) \right] Y^{}_{\rm u} \; , \nonumber \\
16\pi^2 \frac{{\rm d}Y^{}_{\rm d}}{{\rm d}t} &=& \left[ \alpha^{\rm
SM}_{\rm d} + C^{\rm SM}_{\rm d,u} \left(Y^{}_{\rm u} Y^\dagger_{\rm
u} \right) + C^{\rm SM}_{\rm d,d} \left(Y^{}_{\rm d} Y^\dagger_{\rm
d}\right) \right] Y^{}_{\rm d} \; .
\end{eqnarray}
The relevant coefficients in Eqs.~(3) and (A1) are $C^{\rm
SM}_\kappa = C^{\rm SM}_{\rm u,d} = C^{\rm SM}_{\rm d,u} = -3/2$,
$C^{\rm SM}_{l,l} = C^{\rm SM}_{\rm u,u} = C^{\rm SM}_{\rm d,d} =
+3/2$, and
\begin{eqnarray}
\alpha^{\rm SM}_\kappa &=& - 3g^2_2 + \lambda + 2T^{\rm SM}_{\rm M}
\; , \nonumber \\
\alpha^{\rm SM}_l &=& -\frac{9}{4} g^2_1 - \frac{9}{4} g^2_2 +
T^{\rm SM}_{\rm M} \; , \nonumber \\
\alpha^{\rm SM}_{\rm u} &=& -\frac{17}{20} g^2_1 - \frac{9}{4} g^2_2
- 8 g^2_3 + T^{\rm SM}_{\rm M} \; , \nonumber \\
\alpha^{\rm SM}_{\rm d} &=& -\frac{1}{4} g^2_1 - \frac{9}{4} g^2_2 -
8 g^2_3 + T^{\rm SM}_{\rm M} \;
\end{eqnarray}
with $T^{\rm SM}_{\rm M} \equiv {\rm tr}\left[3\left(Y^{}_{\rm
u}Y^\dagger_{\rm u}\right) + 3\left(Y^{}_{\rm d}Y^\dagger_{\rm
d}\right) + \left(Y^{}_lY^\dagger_l\right)\right]$. The RGE's for
the $SU(3)_{\rm C} \times SU(2)_{\rm L}\times U(1)^{}_{\rm Y}$ gauge
couplings $g^{}_3$, $g^{}_2$, and $g^{}_1$ are given by
\begin{equation}
16\pi^2 \frac{{\rm d}g^{}_i}{{\rm d}t} = b^{\rm SM}_i g^3_i
\end{equation}
with $(b^{\rm SM}_1, b^{\rm SM}_2, b^{\rm SM}_3) = (41/10, -19/6,
-7)$. The quartic coupling $\lambda$ of the Higgs field appears in
the RGE of $\kappa$, which affects the evolution of absolute
neutrino masses. It should satisfy the following RGE
\begin{eqnarray}
16\pi^2 \frac{{\rm d}\lambda}{{\rm d}t} &=& 6\lambda^2 -
3\lambda\left(\frac{3}{5}g^2_1 + 3g^2_2\right) + \frac{3}{2}
\left(\frac{9}{25}g^2_1 + \frac{6}{5}g^2_1 g^2_2 + 3g^2_2\right) \nonumber \\
~ &~& + 4\lambda T^{\rm SM}_{\rm M} - 8~{\rm
tr}\left[3\left(Y^{}_{\rm u}Y^\dagger_{\rm u}\right)^2 +
3\left(Y^{}_{\rm d}Y^\dagger_{\rm d}\right)^2 +
\left(Y^{}_lY^\dagger_l\right)^2\right] \; .
\end{eqnarray}
It is worth mentioning that if the experimental uncertainties of the
top quark mass $M^{}_t$ and the strong coupling $\alpha^{}_s$ are
taken into account, the SM vacuum could be stable up to the Planck
scale $\Lambda^{}_{\rm Pl} = 1.2\times 10^{19}~{\rm GeV}$
\cite{Vacuum}, even for a Higgs mass $M^{}_H = 125~{\rm GeV}$
indicated by the recent results of the ATLAS and CMS experiments.

\subsection{The MSSM}

In the MSSM, the RGE's in Eqs.~(3) and (A1) are still applicable,
but the relevant flavor-universal coefficients are as follows:
$C^{\rm MSSM}_\kappa = C^{\rm MSSM}_{\rm u,d} = C^{\rm MSSM}_{\rm
d,u} = 1$, $C^{\rm MSSM}_{l,l} = C^{\rm MSSM}_{\rm u,u} = C^{\rm
MSSM}_{\rm d,d} = 3$, and
\begin{eqnarray}
\alpha^{\rm MSSM}_\kappa &=& - \frac{6}{5} g^2_1 - 6 g^2_2 + 6~{\rm
tr}\left(Y^{}_{\rm u}Y^\dagger_{\rm u}\right) \; , \nonumber \\
\alpha^{\rm MSSM}_l &=& -\frac{9}{5} g^2_1 - 3 g^2_2 + {\rm
tr}\left[ 3\left(Y^{}_{\rm d}Y^\dagger_{\rm d}\right)
+ \left(Y^{}_l Y^\dagger_l\right)\right] \; , \nonumber \\
\alpha^{\rm MSSM}_{\rm u} &=& -\frac{13}{15} g^2_1 - 3 g^2_2 -
\frac{16}{3} g^2_3 + 36~{\rm tr} \left( Y^{}_{\rm u}Y^\dagger_{\rm
u} \right) \; , \nonumber \\
\alpha^{\rm MSSM}_{\rm d} &=& -\frac{7}{15} g^2_1 - 3g^2_2 -
\frac{16}{3} g^2_3 + {\rm tr}\left[ 3\left(Y^{}_{\rm
d}Y^\dagger_{\rm d}\right) + \left(Y^{}_l Y^\dagger_l\right)\right]
\; .
\end{eqnarray}
The RGE's for the gauge couplings are given in Eq.~(A3), but with
$(b^{\rm MSSM}_1, b^{\rm MSSM}_2, b^{\rm MSSM}_3) = (33/5, 1, -3)$
in the beta functions. As we can see from the RGE of $\kappa$, the
running neutrino parameters are determined by the charged-lepton
Yukawa coupling matrix $Y^{}_l$, especially the tau-lepton Yukawa
coupling $y^2_\tau = m^2_\tau (1 + \tan^2 \beta)/v^2$, which could
significantly be enhanced for a large value of $\tan\beta$. Such a
unique feature can make the RG running of leptonic mixing parameters
remarkable in the MSSM.

\subsection{The UEDM}

In the UEDM, all the SM fields are promoted to a higher-dimensional
spacetime, so every SM particle is accompanied by a tower of
Kaluza--Klein (KK) modes \cite{UED1}. In the simplest UEDM with only
one extra spatial dimension, which is compactified on an
$S^1/Z^{}_2$ orbifold with radius $R$, the KK parity defined as
$(-1)^n$ for the $n$-th KK mode is conserved after compactification.
The mass scale of the first excited KK mode, i.e., $\mu^{}_0 \equiv
R^{-1}$, has been constrained to be larger than about $300~{\rm
GeV}$.

If we extend the UEDM by an effective operator $(\overline{\ell} H)
\cdot \hat{\kappa} \cdot (H^T \ell^C)/2$ to accommodate Majorana
neutrino masses, just as in Eq.~(1), then the effective Majorana
neutrino mass matrix after electroweak symmetry breaking is
$M^{}_\nu = \kappa v^2$ with $\kappa = \hat{\kappa}/(\pi R)$. The
RGE of $\kappa$ now receives contributions from the KK modes, which
are excited at the energy scale of interest. More explicitly, the
RGE's for $\kappa$ and the Yukawa coupling matrices of the charged
fermions are also given by Eqs.~(3) and (A1), but with the following
coefficients \cite{UED1}
\begin{eqnarray}
\alpha^{\rm UEDM}_\kappa &=& \alpha^{\rm SM}_\kappa + s
\left(-\frac{1}{4} g^2_1 - \frac{11}{4} g^2_2 + \lambda + 4 T^{\rm
SM}_{\rm M}\right) \; , \nonumber \\
\alpha^{\rm UEDM}_l &=& \alpha^{\rm SM}_l + s \left(-\frac{33}{8}
g^2_1 - \frac{15}{8} g^2_2 + 2T^{\rm SM}_{\rm M}\right) \; ,
\nonumber \\
\alpha^{\rm UEDM}_{\rm u} &=& \alpha^{\rm SM}_{\rm u} + s
\left(-\frac{101}{72} g^2_1 - \frac{15}{8} g^2_2 - \frac{28}{3}
g^2_3 + 2T^{\rm SM}_{\rm M}\right) \; , \nonumber \\
\alpha^{\rm UEDM}_{\rm d} &=& \alpha^{\rm SM}_{\rm d} + s
\left(-\frac{17}{72} g^2_1 - \frac{15}{8} g^2_2 - \frac{28}{3} g^2_3
+ 2T^{\rm SM}_{\rm M}\right) \; ,
\end{eqnarray}
and $C^{\rm UEDM}_{\rm x} = C^{\rm SM}_{\rm x} (1+s)$ with ``$\rm
x$" being any relevant subscript. Note that $s \equiv \lfloor
\mu/\mu^{}_0 \rfloor$ counts the number of excited KK modes for a
given energy scale $\mu$. In addition, the coefficients in the beta
functions of gauge couplings turn out to be
\begin{eqnarray}
b^{\rm UEDM}_1 = b^{\rm SM}_1 + \frac{27}{2} s \; , ~~~ b^{\rm
UEDM}_2 = b^{\rm SM}_2 + \frac{7}{6} s \; , ~~~ b^{\rm UEDM}_3 =
b^{\rm SM}_3 -\frac{5}{2} s \; .
\end{eqnarray}
Finally, the RGE for the quartic Higgs coupling $\lambda$ is quite
relevant in the UEDM, as in the SM case. It has been found to be
\cite{UED1}
\begin{eqnarray}
16\pi^2 \frac{{\rm d}\lambda}{{\rm d}t} &=& 6(1+s)\lambda^2 -
3(1+s)\lambda\left(\frac{3}{5}g^2_1 + 3g^2_2\right) + \frac{3}{2}
(1+\frac{4}{3}s) \left(\frac{9}{25}g^4_1 + \frac{6}{5} g^2_1 g^2_2 +
3 g^2_4\right) \nonumber
\\ &~&
+ 4 (1+2s)\lambda T^{\rm SM}_{\rm M} - 8(1+2s)~{\rm
tr}\left[3\left(Y^{}_{\rm u}Y^\dagger_{\rm u}\right)^2 +
3\left(Y^{}_{\rm d}Y^\dagger_{\rm d}\right)^2 +
\left(Y^{}_lY^\dagger_l\right)^2\right] \; .
\end{eqnarray}

\section{RGE's for Dirac neutrinos}

If the SM is extended with three right-handed neutrino singlets,
then neutrinos acquire Dirac masses in the same way as the charged
leptons and quarks do. At one-loop level, the RGE's of the fermion
Yukawa coupling matrices read \cite{LindnerD}
\begin{eqnarray}
16\pi^2 \frac{{\rm d}Y^{}_\nu}{{\rm d}t} &=& \left[\alpha^{\rm
SM}_\nu + C^{\rm SM}_{\nu,\nu} \left(Y^{}_\nu Y^\dagger_\nu\right) +
C^{\rm SM}_{\nu, l} \left(Y^{}_l Y^\dagger_l\right) \right] Y^{}_\nu
\; , \nonumber \\
16\pi^2 \frac{{\rm d}Y^{}_l}{{\rm d}t} &=& \left[\alpha^{\rm SM}_l +
C^{\rm SM}_{l,\nu} \left(Y^{}_\nu Y^\dagger_\nu\right) + C^{\rm
SM}_{l,l} \left(Y^{}_l Y^\dagger_l\right)\right] Y^{}_l \; ,
\nonumber \\
16\pi^2 \frac{{\rm d}Y^{}_{\rm u}}{{\rm d}t} &=& \left[ \alpha^{\rm
SM}_{\rm u} + C^{\rm SM}_{\rm u,u} \left(Y^{}_{\rm u} Y^\dagger_{\rm
u} \right) + C^{\rm SM}_{\rm u,d} \left(Y^{}_{\rm d} Y^\dagger_{\rm
d}\right) \right] Y^{}_{\rm u} \; , \nonumber \\
16\pi^2 \frac{{\rm d}Y^{}_{\rm d}}{{\rm d}t} &=& \left[ \alpha^{\rm
SM}_{\rm d} + C^{\rm SM}_{\rm d,u} \left(Y^{}_{\rm u} Y^\dagger_{\rm
u} \right) + C^{\rm SM}_{\rm d,d} \left(Y^{}_{\rm d} Y^\dagger_{\rm
d}\right) \right] Y^{}_{\rm d} \; ,
\end{eqnarray}
where $C^{\rm SM}_{\rm f, g} = +3/2$ (for ${\rm f} = {\rm g}$) and
$-3/2$ (for ${\rm f}\neq {\rm g}$), and
\begin{eqnarray}
\alpha^{\rm SM}_\nu &=& - \frac{9}{20} g^2_1 - \frac{9}{4} g^2_2 +
T^{\rm SM}_{\rm D}
\; , \nonumber \\
\alpha^{\rm SM}_l &=& -\frac{9}{4} g^2_1 - \frac{9}{4} g^2_2 +
T^{\rm SM}_{\rm D} \; , \nonumber \\
\alpha^{\rm SM}_{\rm u} &=& -\frac{17}{20} g^2_1 - \frac{9}{4} g^2_2
- 8 g^2_3 + T^{\rm SM}_{\rm D} \; , \nonumber \\
\alpha^{\rm SM}_{\rm d} &=& -\frac{1}{4} g^2_1 - \frac{9}{4} g^2_2 -
8 g^2_3 + T^{\rm SM}_{\rm D} \;
\end{eqnarray}
with $T^{\rm SM}_{\rm D} \equiv {\rm tr}\left[3\left(Y^{}_{\rm
u}Y^\dagger_{\rm u}\right) + 3\left(Y^{}_{\rm d}Y^\dagger_{\rm
d}\right) + \left(Y^{}_\nu Y^\dagger_\nu\right) +
\left(Y^{}_lY^\dagger_l\right)\right]$. The RGE's of fermion Yukawa
coupling matrices are the same as in Eq.~(B1) for the MSSM, but with
different coefficients, namely $C^{\rm MSSM}_{\rm f, g} = +3$ (for
${\rm f} = {\rm g}$) and $+1$ (for ${\rm f}\neq {\rm g}$), and
\begin{eqnarray}
\alpha^{\rm MSSM}_\nu &=& - \frac{3}{5} g^2_1 - 3 g^2_2 + {\rm
tr}\left[3\left(Y^{}_{\rm u} Y^\dagger_{\rm u}\right) +
\left(Y^{}_\nu Y^\dagger_\nu\right)\right]
\; , \nonumber \\
\alpha^{\rm MSSM}_l &=& -\frac{9}{5} g^2_1 - 3 g^2_2 + {\rm
tr}\left[3\left(Y^{}_l Y^\dagger_l\right) +
\left(Y^{}_\nu Y^\dagger_\nu\right)\right] \; , \nonumber \\
\alpha^{\rm MSSM}_{\rm u} &=& -\frac{13}{15} g^2_1 - 3 g^2_2 -
\frac{16}{3} g^2_3 + {\rm tr}\left[3\left(Y^{}_{\rm u}
Y^\dagger_{\rm u}\right) +
\left(Y^{}_\nu Y^\dagger_\nu\right)\right] \; , \nonumber \\
\alpha^{\rm MSSM}_{\rm d} &=& -\frac{7}{15} g^2_1 - 3 g^2_2 -
\frac{16}{3} g^2_3 + {\rm tr}\left[3\left(Y^{}_{\rm d}
Y^\dagger_{\rm d}\right) + \left(Y^{}_l Y^\dagger_l\right)\right]
\;.
\end{eqnarray}
The RGE's of three gauge couplings $g^{}_1$, $g^{}_2$, and $g^{}_3$
are the same as those in the case of Majorana neutrinos [see
Eq.~(A3)].


\end{document}